\documentclass[12pt]{article}
\usepackage{latexsym}
\usepackage{amsmath,amssymb}
\usepackage{bm}
\usepackage[dvipdfmx]{graphicx,color}
\usepackage{amssymb}
\usepackage{cite}
\usepackage{amsfonts, amscd, amsthm}
\usepackage[all]{xy}
\usepackage{here}
\usepackage{ulem}
\usepackage{mathabx}

\newcommand{\sn}{\rm sn}

\renewcommand{\theequation}{\thesection.\arabic{equation}}

\newcommand{\Pring}{\mathrm{\mathring{\wp}}}
\newcommand{\dd}{{\rm d}}

\makeatletter
\@addtoreset{equation}{section}
\renewcommand{\theequation}{\thesection.\@arabic\c@equation}
\makeatother

\makeatletter
\renewcommand\appendix{\par%\newpage
  \setcounter{section}{0}%
  \setcounter{subsection}{0}%
  \gdef\thesection{Appendix \@Alph\c@section }
  \renewcommand{\theequation}
  {\Alph{section}.\arabic{equation}}
}
\makeatother

\makeatletter
\newcounter{subeqncnt}
\def\thesubeqncnt{\alph{subeqncnt}}
\def\subequations{\begingroup%
\stepcounter{equation}\edef\@tempa{\theequation}%
\let\c@equation\c@subeqncnt\c@subeqncnt\z@
\edef\theequation{\@tempa\noexpand\thesubeqncnt}}

\makeatother
\allowdisplaybreaks

\begin{document}

\titlepage

%%%%%
\title{Differential Equations \\of Genus Four 
Hyperelliptic $\wp$ Functions\\
} 
\author{Masahito Hayashi\thanks{masahito.hayashi@oit.ac.jp}\\
Osaka Institute of Technology, Osaka 535-8585, Japan\\
Kazuyasu Shigemoto\thanks{shigemot@tezukayama-u.ac.jp} \\
Tezukayama University, Nara 631-8501, Japan\\
Takuya Tsukioka\thanks{tsukioka@bukkyo-u.ac.jp}\\
Bukkyo University, Kyoto 603-8301, Japan\\
}
\date{\empty}

%%%%%%%%%%%%%%%%%%%%%%%%%%%%%%%%%%%%%%%%

\maketitle
\abstract{%
In order to find higher dimensional integrable models, we study differential equations of hyperelliptic $\wp$ functions up to genus four. For genus two, differential equations of hyperelliptic $\wp$ functions can be written in the Hirota form. If the genus is more than one, we have KdV equation. If the genus is more than two, we have KdV and another KdV equations. If the genus becomes more than three, there appear differential equations which cannot be written in the Hirota form, which means that the Hirota form is not enough to characterize the integrable differential equations. We have shown that some differential equations are satisfied for general genus. We can obtain differential equations for general genus step by step. 
}

%%%%%%%%%%%%%%%%%%%%%%%%%%%%%%%%%%%%%%%%%%%%%%%%%
\section{Introduction} 
\setcounter{equation}{0}

Through studies of soliton system, we have solved non-linear problems of very interesting phenomena. Starting from the inverse scattering method~\cite{Gardner,Lax,Zakhrov}, many interesting developments have been done including the AKNS formulation~\cite{Ablowitz}, 
the geomertical approach and various 
developments~\cite{Bianchi,Hermann,Sasaki,Chern1,Chern2,Reyes},
the B\"{a}cklund transformation~\cite{Wahlquist,Wadati1,Wadati2}, the Hirota equation~\cite{Hirota1,Hirota2}, the Sato theory~\cite{Sato}, the vertex construction of the soliton solution~\cite{Miwa1,Date1,Jimbo1}, and the Schwarzian type mKdV/KdV equation~\cite{Weiss}. Soliton theory is, in some sense, the prototype of the superstring theory, because the M\"{o}bius transformation, vertex construction and AdS structure are used to understand the structure of soliton system. Our understanding of the soliton is still in progress. 

In our previous papers, 
by using the Lie group structure as our guiding principle,
we have revealed that the two dimensional integrable models such as KdV/mKdV/sinh-Gordon are the consequence of the SO(2,1)$\cong$ SL(2,$\mathbb{R}$) Lie group structure~\cite{Hayashi1,Hayashi2,Hayashi3,Hayashi4,Hayashi5}
\footnote{
 The relation between the KdV equation and SL(2,$\mathbb{R}$) Lie group suructure goes back to
Hermann's paper in 1976~\cite{Hermann}.  The relation between the sine-Gordon equation and 
SO(2,1) Lie group suructure goes back to Bianchi's paper in 1879~\cite{Bianchi}
}
%.
Here we would like to to study higher-dimensional integrable models. KdV/mKdV/sinh-Gordon equations and KP equations are typically understood as two- and three-dimensional integrable models, respectively. First, we would like to know whether there exists a universality of the integrable models, that is, whether any two- and three-dimensional integrable models always contain KdV/mKdV/sinh-Gordon equations and KP equations, respectively. 

For higher-dimensional integrable models, there is a soliton type approach of Kyoto school ~\cite{Sato,Miwa1,Date1,Jimbo1} where they use the special fermion, which generates $N$-soliton solutions. Starting with the fermionic bilinear identity of $\mathfrak{gl}(\infty, \mathbb{R})$, they have obtained KP hierarchy and finite higher-dimensional Hirota forms by the reduction of KP hierarchy. Another systematic approach to high-dimensional integrable models is to find differential equations for higher genus hyperelliptic functions by using the analogy of differential equation of Weierstrass $\wp$ function. By solving the Jacobi's inversion problem, the integrability of hyperelliptic functions are automatically guaranteed, since the integrability condition and the single-valuedness are equivalent for hyperelliptic functions. So far, only for genus one, two~\cite{Baker3} and three~\cite{Baker4,Buchstaber1,Buchstaber2} cases are studied because it becomes difficult to solve the Jacobi's inversion problem and obtain differential equations for higher genus cases. 

In 1981, Dubrobin found the hyperelliptic solution of the KdV or the KP equation for genus two and three cases~\cite{Dubrovin}. While, we want to find what kinds of and how many Weierstrass type differential equations such as $\textrm{d}^2 \wp(x)/\textrm{d}x^2=6 \wp^2(x)-g_2/2$ for various genus are there, which give us the clue to understand the Lie group structure of hyperelliptic $\wp$ functions. Matsutani pointed out in 2001~\cite{Matsutani}, that Baker had already found hyperelliptic solution for KdV and KP equation for genus three case in 1897~\cite{Baker4}.

In this paper, we study to obtain differential equations of genus four case. In the approach, we would like to examine the connections between i) higher-dimensional integrable differential equations, ii) higher-rank Lie group structure and iii) higher genus hyperelliptic functions.

%%%%%%%%%%%%%%%%%%%%%%%%%%%%%%%%%%%%%%%%%%%%%%%%%%%%%%%%%%%
\section{Formulation of Differential Equations in General Genus and the Review of Genus Two and Three Cases} 
\setcounter{equation}{0}

%%%%%%%%%%%%%%%%%%%%%%
\subsection{Formulation of differential equations in general genus}

We summarize the formulation of hyperelliptic $\wp$ function according to Baker's work~\cite{Baker1,Baker2,Baker3,Baker4}. We consider the genus $g$ hyperelliptic curve
\begin{equation}
C:\quad y_i^2=\sum_{k=0}^{2g+2} \lambda_k x_i^k, \qquad  (i=1, 2, \cdots, g). 
\label{2e1}
\end{equation}
The Jacobi's inversion problem consists of solving the following system
\begin{equation}
\dd u_1    =\sum_{i=1}^g \frac{         \dd x_i}{y_i}, \quad 
\dd u_2    =\sum_{i=1}^g \frac{x_i      \dd x_i}{y_i}, \quad \cdots,\quad  
\dd u_{g-1}=\sum_{i=1}^g \frac{x_i^{g-2}\dd x_i}{y_i}, \quad 
\dd u_g    =\sum_{i=1}^g \frac{x_i^{g-1}\dd x_i}{y_i}.
\label{2e2}
\end{equation} 
From these equations, we have 
\begin{equation}
\frac{\partial x_i}{\partial u_j}=\frac{y_i \chi_{g-j}\left(x_i; x_1,x_2,\cdots, x_g\right)}{F'(x_i)},
\label{2e3}
\end{equation}
by using the relation
\begin{equation}
\sum_{i=1}^g \frac{x_i^{k-1} \chi_{g-j}(x_i; x_1, x_2, \cdots, x_g)}{F'(x_i)}
=\delta_{k j},\quad (1\leq j \leq g).
\label{2e4}
\end{equation}
We define
$\displaystyle{F(x)=\prod_{i=1}^g (x-x_i)}$ and denote
$F'(x_i)$ as $\displaystyle{F'(x_i)=\frac{\dd F(x)}{\dd x}\Big|_{x=x_i}}$. 
For example,
$F'(x_1)=(x_1-x_2)(x_1-x_3)\cdots (x_1-x_g)$ .
For $\chi_{g-j}(x_i; x_1,x_2,\cdots, x_g)$, we first define the following generalized function
\begin{align}
\chi_{g-j}(x;x_1, \cdots, x_p) = &\ x^{g-j}-h_1(x_1, \cdots, x_p) x^{g-j-1} \nonumber\\
&+h_2(x_1, x_2, \cdots, x_p) x^{g-j-2}+\cdots+(-1)^{g-j} h_{g-j}(x_1, \cdots, x_p)  ,
\label{2e5}
\end{align}
where $h_j(x_1, \cdots, x_p)$ is the $j$-th fundamental symmetric polynomial basis of 
$\{x_1, \cdots, x_p\}$, i.e.\, 
\begin{equation}
\prod_{i=1}^p (x-x_i)=x^p+\sum_{j=1}^p (-1)^j h_j(x_1,x_2,\cdots, x_p) x^{p-j}.
\label{2e6}
\end{equation}
Putting $p=g$ and $x=x_k$ in $\chi_{g-j}(x;x_1,x_2,\cdots, x_p)$, we have $\chi_{g-j}(x_i; x_1,x_2,\cdots, x_g)$ in  the following form
\begin{align}
\chi_{g-j}(x_i;x_1,x_2,\cdots, x_g) 
=&\ x_i^{g-j}-h_1(x_1,x_2,\cdots, x_g) x_i^{g-j-1} \nonumber\\
&+h_2(x_1,x_2,\cdots, x_g) x_i^{g-j-2}
+\cdots+(-1)^{g-j} h_{g-j}(x_1,x_2,\cdots, x_g)   .
\label{2e7}
\end{align}
For example 
\begin{align*}
\chi_0(x_1;x_1,x_2,\cdots, x_g)&=1, \nonumber\\
\chi_1(x_1;x_1,x_2,\cdots, x_g)&=x_1  -(x_1+x_2+\cdots+x_g)=-h_1(x_2, x_3,\cdots,x_g), \nonumber\\
\chi_2(x_1;x_1,x_2,\cdots, x_g)&=x_1^2-(x_1+x_2+\cdots+x_g)x_1+(x_1 x_2+x_1 x_3+\cdots)\nonumber\\
                               &=x_2 x_3+x_2 x_4+\cdots=h_2(x_2, x_3,\cdots,x_g),      \nonumber\\
                               &~\:\vdots 
\label{2e8}
\end{align*}
From Eq.(\ref{2e6}), we have
\begin{equation}
x_i^g    -h_1(x_1,x_2,\cdots,x_g)x_i^{g-1}
         +h_2(x_1,x_2,\cdots,x_g)x_i^{g-2}
+\cdots
+(-1)^{g} h_g(x_1,x_2,\cdots,x_g)=0 .
\label{2e9}
\end{equation}
The $\zeta_j$ functions are given from the hyperelliptic curve in the following way~\cite{Baker1} 
\begin{equation}
\dd(-\zeta_j)=
\sum_{i=1}^g\frac{\dd x_i}{y_i}
\sum_{k=j}^{2g+1-j} (k+1-j) \lambda_{k+1+j} x_i^k
-2\dd\left(\sum_{i=1}^g\frac{y_i \chi_{g-j-1}(x_i; x_1, \cdots, \widecheck{x}_i, \cdots,x_g)}{F'(x_i)}\right),
\label{2e10}
\end{equation} 
where $\widecheck{x}_j$ denotes that the $x_j$ variable is missing. The overall factor of $\zeta_j$ is different  from  the textbook of Baker~\cite{Baker1} and papers of Buchstaber {\it et al.}~\cite{Buchstaber1,Buchstaber2}. We changed it to make it easier to see the dual symmetry of differential equations in $\S 4.1$. In this expression, we can show $\dd(-\zeta_0)=0$ in the following way
\begin{align}
\dd(-\zeta_0)
&=\sum_{i=1}^g \frac{\dd x_i}{y_i}\sum_{k=0}^{2g+1} (k+1) \lambda_{k+1} x_i^k
-2\dd\left(\sum_{i=1}^g\frac{y_i \chi_{g-1}(x_i; x_1, \cdots,
\widecheck{x}_i, \cdots,x_g)}{F'(x_i)}\right)
\nonumber\\
&=\sum_{i=1}^g 
\frac{1}{y_i}\dd\left(\sum_{l=0}^{2g+2} \lambda_{l} x_i^{l}\right)
-2\dd\left(\sum_{i=1}^g y_i \right)
=\sum_{i=1}^g
\frac{1}{y_i}\dd\left(y_i^2 \right)-2\dd \left(\sum_{i=1}^g y_i \right)
\nonumber 
\\
&=0,  
\label{2e11}
\end{align} 
where we use $\chi_{g-1}(x_i; x_1, x_2, \cdots, \widecheck{x}_i, \cdots,x_g)=F'(x_i)$. In addition, an identity $\dd(-\zeta_{g+1})=0$ can be derived easily as follows
\begin{align}
\dd(-\zeta_{g+1})
&=\sum_{i=1}^g \frac{\dd x_i}{y_i}\sum_{k=g+1}^{g} (k-g) \lambda_{k+g+2} x_i^k
-2\dd\left(\sum_{i=1}^g\frac{y_i \chi_{-2}(x_i; x_1, \cdots,
\widecheck{x}_i, \cdots,x_g)}{F'(x_i)}\right)
\nonumber
\\
&=0,  
\label{2e11add}
\end{align} 
because there is no such sum of $k$ and we have $\chi_{-2}=0$. These $\zeta_j(u_1,u_2,\cdots, u_g)$ satisfy the integrability condition 
\begin{equation}
 \frac{\partial \left(-\zeta_j(u_1,u_2,\cdots, u_g)\right)}{\partial u_k}
=\frac{\partial \left(-\zeta_k(u_1,u_2,\cdots, u_g)\right)}{\partial u_j}  .
\label{2e12}
\end{equation} 
In the Baker's textbook~\cite{Baker1}, the expression of the second term of the r.h.s of Eq.(\ref{2e10}) is misleading.%
%
%%%%%%%%%%%%%%%%%%%%
\footnote{%
In the 1897 edition of the textbook of Baker~\cite{Baker1}, the expression of p.320 in \S 214  is correct, but the expression of p.321 in \S 215 is wrong. If we use the exprssion of p.321 in \S 215, the integrability condition Eq.(\ref{2e12}) is not satisfied.}
%%%%%%%%%%%%%%%%%%%%
%
$\wp_{jk}(u_1,u_2,\cdots, u_g)$ functions are given from the above $\zeta_j(u_1,u_2,\cdots, u_g)$ functions in the form 
\begin{equation}
\wp_{jk}(u_1,u_2,\cdots, u_g)=\wp_{kj}(u_1,u_2,\cdots, u_g)
=\dfrac{\partial \left(-\zeta_j(u_1,u_2,\cdots, u_g)\right)}{\partial u_k}.
\label{2e13}
\end{equation}
By using Eq.(\ref{2e11}) and Eq.(\ref{2e11add}), we have
\begin{equation}
\wp_{k0}=\wp_{k,g+1}=0, \quad (1\le k \le g).
\label{2e13add}
\end{equation}
These $\zeta_j$, $\wp_{jk}$ and $\wp_{j k l m}$ are given by the hyperelliptic $\sigma$ function in the form%
%
%%%%%%%%%%%%%%%%%%%%
\footnote{%
We do not use the explicit form of the $\sigma$ function later, but we give the expression of $\zeta_j$, $\wp_{jk}$ and $\wp_{j k l m}$ using the $\sigma$ function just only for the explanation of the index $i, jk, jklm$. For the definition of the $\sigma$ function, see for example papers of Buchstaber {\it et al.}~\cite{Buchstaber1,Buchstaber2}.}
%%%%%%%%%%%%%%%%%%%%
%
$$
-\zeta_j=\dfrac{\partial (-\log \sigma )}{\partial u_j}, \quad  
\wp_{jk}=\dfrac{\partial^2 (-\log \sigma )}{\partial u_j \partial u_k}, \quad  
\textrm{and} \quad 
\wp_{j k l m}=\dfrac{\partial^4(-\log \sigma )}
{\partial u_j \partial u_k \partial u_l \partial u_m}, \qquad \textrm{etc.}.
$$

For the Weierstrass type, i.e.\,$\lambda_{2g+2}=0$, we have $\displaystyle{\dd(-\zeta_g)=\lambda_{2g+1}\sum_{i=1}^{g} \frac{x_i^g\dd x_i }{y_i}}$, which gives
\begin{align}
\widehat{\wp}_{g g  }(u_1,u_2,\cdots, u_g)
 &=\frac{1}{\lambda_{2g+1}}\wp_{g g  }(u_1,u_2,\cdots, u_g)= h_1(x_1,x_2,\cdots,x_g), 
\label{2e14}\\
\widehat{\wp}_{g,g-1}(u_1,u_2,\cdots, u_g)
 &=\frac{1}{\lambda_{2g+1}}\wp_{g,g-1}(u_1,u_2,\cdots, u_g)=-h_2(x_1,x_2,\cdots,x_g), 
\label{2e15}\\
 &\ \vdots \nonumber\\
\widehat{\wp}_{g 1  }(u_1,u_2,\cdots, u_g)
 &=\frac{1}{\lambda_{2g+1}}\wp_{g 1  }(u_1,u_2,\cdots, u_g)=(-1)^{g-1} h_g(x_1,x_2,\cdots,x_g), 
\label{2e16}
\end{align}
by using 
\begin{equation}
\sum_{i=1}^g \frac{x_i^{g} \chi_{g-j}(x_i;x_1, x_2, \cdots, x_g)}{F'(x_i)} 
=(-1)^{g-j} h_{g-j+1}(x_1, x_2,\cdots,x_g)   .
\label{2e17}
\end{equation}
Then we have 
\begin{equation}
x_i^g=\sum_{j=1}^{g} \widehat{\wp}_{g j} x_i^{j-1} 
     =\widehat{\wp}_{gg   } x_i^{g-1}
     +\widehat{\wp}_{g,g-1} x_i^{g-2}
     +\cdots
     +\widehat{\wp}_{g2} x_i
     +\widehat{\wp}_{g1}.
\label{2e18}
\end{equation}
We can easily show Eq.(\ref{2e4}) and Eq.(\ref{2e17}) by using Eq.(\ref{2e7}), Eq.(\ref{2e9}) and the following relation~\cite{Shigemoto2}
\begin{equation}
\sum_{i=1}^g \frac{x_i^{j-1} }{F'(x_i)}=\delta_{j g},\qquad (1\leq j \leq g) .
\label{2e19}
\end{equation}
In this way, we have $\displaystyle{\dd(-\zeta_g)=\sum_{j=1}^{g} \wp_{g j}\dd u_j}$. For other $\wp_{i j}$, we must use $\zeta_j$, which satisfies the integrability condition Eq.(\ref{2e12}).  

%%%%%%%%%%%%%%%%%%
\subsection{Differential equations of genus two hyperelliptic $\wp$ functions} 

We here review the genus two hyperelliptic $\wp$ function. The hyperelliptic curve in this case is given by 
\begin{equation}
C:\quad y_i^2 = \lambda_6 x_i^6 + \lambda_5 x_i^5 + \lambda_4 x_i^4 + \lambda_3 x_i^3
              + \lambda_2 x_i^2 + \lambda_1 x_i   + \lambda_0 , \quad (i=1, 2).
\label{2e20}
\end{equation}
The Jacobi's inversion problem consists of solving the following system
\begin{equation}
\dd u_1=\frac{    \dd x_1}{y_1}+\frac{    \dd x_2}{y_2}, \qquad 
\dd u_2=\frac{x_1 \dd x_1}{y_1}+\frac{x_2 \dd x_2}{y_2}   . 
\label{2e21}
\end{equation} 
Then we have 
\begin{equation}
\frac{\partial x_1}{\partial u_2}= \frac{y_1    }{x_1-x_2}, \qquad 
\frac{\partial x_2}{\partial u_2}=-\frac{y_2    }{x_1-x_2}, \qquad 
\frac{\partial x_1}{\partial u_1}=-\frac{x_2 y_1}{x_1-x_2}, \qquad 
\frac{\partial x_2}{\partial u_1}= \frac{x_1 y_2}{x_1-x_2}.
\label{2e22}
\end{equation}
In this case,
\begin{align}
\dd(-\zeta_2)
 &=\sum_{i=1}^2 \frac{ \left(2 \lambda_6 x_i^3+ \lambda_5 x_i^2 \right)\dd x_i}{y_i}  ,
\label{2e23}\\
\dd(-\zeta_1)
 &=\sum_{i=1}^2 \frac{ \left(4 \lambda_6 x_i^4+3\lambda_5 x_i^3+2 \lambda_4 x_i^2+\lambda_3 x_i\right)\dd x_i}{y_i}
  -2\dd\left(\frac{y_1-y_2}{x_1-x_2}\right)  .
\label{2e24}
\end{align}
For these $\zeta_{1}$ and $\zeta_{2}$, we have checked the integrability condition $\partial \zeta_1/\partial u_2=\partial \zeta_2/\partial u_1$. We use the useful functions $\widehat{\wp}_{22}$, $\widehat{\wp}_{21}$ and $\widehat{\wp}_{11}$ of the form
\begin{align}
\widehat{\wp}_{22}
 &=\frac{1}{\lambda_5} \wp_{22}
  =\frac{1}{\lambda_5}\frac{\partial (- \zeta_2)}{\partial u_2}
  =x_1+x_2+\frac{2 \lambda_6}{\lambda_5}\left(x_1^2+x_1 x_2+x_2^2\right)  ,
\label{2e25}\\
\widehat{\wp}_{21}
 &=\frac{1}{\lambda_5} \wp_{21}
  =\frac{1}{\lambda_5} \frac{\partial (- \zeta_2)}{\partial u_1}
  =-x_1 x_2-\frac{2 \lambda_6}{\lambda_5} x_1 x_2 \left(x_1+x_2\right)  ,
\label{2e26}\\
\widehat{\wp}_{11}
 &=\frac{1}{\lambda_5} \wp_{11}
  =\frac{1}{\lambda_5} \frac{\partial (- \zeta_1)}{\partial u_1}
  =\frac{1}{\lambda_5} \frac{F(x_1,x_2)-2 y_1 y_2}{(x_1-x_2)^2}
  +\frac{2 \lambda_6}{\lambda_5} x_1 ^2 x_2^2  ,
\label{2e27}
\end{align}
where 
\begin{align*}
F(x_1,x_2)
=&2 \lambda_6 x_1^3 x_2^3+\lambda_5 x_1^2 x_2^2 (x_1+x_2)+2 \lambda_4 x_1^2 x_2^2
\\ 
 & +\lambda_3 x_1 x_2 (x_1+x_2)+2 \lambda_2 x_1 x_2+\lambda_1 (x_1+x_2)+2\lambda_0. 
\end{align*}
Defining $\Pring_{22}=x_1+x_2, \Pring_{21}=-x_1 x_2$, we have 
\begin{align}
\widehat{\wp}_{22}&=\Pring_{22}+\frac{2 \lambda_6}{\lambda_5}(\Pring_{22}^2+\Pring_{21})  ,
\label{2e28}\\
\widehat{\wp}_{21}&=\Pring_{21}+\frac{2 \lambda_6}{\lambda_5}\Pring_{21} \Pring_{22}  .
\label{2e29}
\end{align}
Then we can express $\Pring_{22}$ and $\Pring_{21}$ as infinite power series of $\widehat{\wp}_{22}$ and $\widehat{\wp}_{21}$. By using Eqs.(\ref{2e22}), (\ref{2e25})--(\ref{2e29}) and with the help of REDUCE, we have the differential equation for $\widehat{\wp}_{22}$ in the form 
\begin{align}
\frac{\partial^2 \widehat{\wp}_{22}}{\partial u_2^2}
  =&\frac{3}{2}\lambda_5 \widehat{\wp}_{22}^2
              +\lambda_4 \widehat{\wp}_{22}
              +\lambda_5 \widehat{\wp}_{21}
            +3 \lambda_6 \widehat{\wp}_{11} 
   +\frac{1}{2}\lambda_3 
\nonumber \\
  &+\frac{2 \lambda_6}{\lambda_5}
      \left( \lambda_6 \left(3 \Pring_{22}^4+6 \Pring_{22} ^2 \Pring_{21}-3\Pring_{21}^2 \right)
            +\lambda_5 \left(3 \Pring_{22}^3+3 \Pring_{22}    \Pring_{21}                \right)
          +3 \lambda_4 \Pring_{22}^2
          +3 \lambda_3 \Pring_{22}
          +2 \lambda_2 \right)  .
\label{2e30}
\end{align}
In order that the differential equation becomes the polynomial type of $\widehat{\wp}_{22}$ and $\widehat{\wp}_{21}$ but not infinite series of these, we must put $\lambda_6=0$
%
%%%%%%%%%%%%%%%%%%%%
\footnote{%
For the general Jacobi type hyperelliptic curve  with $\lambda_{2g+2} \ne 0$, the polynomial type differential equations is not for the hyperelliptic generalization of Weierstrass $\wp$ function of the form $\textrm{d}^2 \wp(x)/\textrm{d}x^2=6 \wp^2(x)-g_2/2$, but for those of Jacobi's $\sn$ function of the form $\textrm{d}^2 \textrm{sn}(x)/\textrm{d}x^2=-(1+k^2) \textrm{sn}(x)+2 k^2 \textrm{sn}^3(x)$, which corresponds to the mKdV equation.}%
%%%%%%%%%%%%%%%%%%%%
%

Then we have 
\begin{align}
\frac{1}{\lambda_5}\wp_{22}
 &=\widehat{\wp}_{22}=\Pring_{22}=x_1+x_2  ,
\label{2e31}\\
\frac{1}{\lambda_5}\wp_{21}
 &=\widehat{\wp}_{21}=\Pring_{21}=-x_1 x_2  ,
\label{2e32}\\
\frac{1}{\lambda_5} \wp_{11}
 &=\widehat{\wp}_{11}
=\frac{1}{\lambda_5} \frac{F(x_1,x_2)|_{\lambda_6=0}-2 y_1 y_2}{(x_1-x_2)^2}  .
\label{2e33}
\end{align}
By using the analogy of the differential equation of Weierstrass $\wp$ function in the form $\textrm{d}^2 \wp(x)/\textrm{d}x^2=6 \wp^2(x)-g_2/2$, with the help of REDUCE, we have the following differential equations~\cite{Baker3}
\begin{align}
1)\quad & \wp_{2222}-\frac{3}{2} \wp_{22}^2
  =\lambda_5 \wp_{21}+\lambda_4\wp_{22}+\frac{1}{2} \lambda_5 \lambda_3  ,
\label{2e34}\\
2)\quad & \wp_{2221}-\frac{3}{2}\wp_{22} \wp_{21}
  =-\frac{1}{2}\lambda_5 \wp_{11}+\lambda_4\wp_{21}  ,
\label{2e35}\\
3)\quad & \wp_{2211}-\wp_{21}^2-\frac{1}{2}\wp_{22} \wp_{11}
  =\frac{1}{2}\lambda_3 \wp_{21}  ,
\label{2e36}\\
4)\quad & \wp_{2111}-\frac{3}{2}\wp_{21} \wp_{11}
  =\lambda_2 \wp_{21}-\frac{1}{2}\lambda_1\wp_{22} -\lambda_5 \lambda_0  ,
\label{2e37}\\
5)\quad & \wp_{1111}-\frac{3}{2}\wp_{11}^2
  =\lambda_2 \wp_{11}+\lambda_1 \wp_{21}-3\lambda_0 \wp_{22}
+\frac{1}{2}\lambda_3 \lambda_1 -2 \lambda_4 \lambda_0  .
\label{2e38}
\end{align}

In addition to $\lambda_6=0$, which is necessary to obtain differential equations of polynomial type, we can always put $\lambda_0=0$ by the constant shift of $x_i$ in Eq.(\ref{2e20}), i.e.\, $x_i \rightarrow x_i+a$ with $\sum_{j=0}^5 \lambda_j a^j=0$. Then, in the standard form of $\lambda_0=0$,  we have some dual symmetry
Eq.(\ref{2e34}) $\leftrightarrow$ Eq.(\ref{2e38}),
Eq.(\ref{2e35}) $\leftrightarrow$ Eq.(\ref{2e37}) and
Eq.(\ref{2e36}) $\leftrightarrow$ Eq.(\ref{2e36})
under $\dd u_2 \leftrightarrow \pm \dd u_1$, $\lambda_1 \leftrightarrow \lambda_5$, 
$\lambda_2 \leftrightarrow \lambda_4$ and 
$\lambda_3 \leftrightarrow \lambda_3$. 

If we differentiate Eq.(\ref{2e34}) with $u_2$, and identify $\wp_{22}(u_1,u_2) \rightarrow u(x,t)$, $\dd u_2\rightarrow \dd x$ and  
$\dd u_1\rightarrow \dd t$, we have
\begin{equation}
u_{xxx}-3 u u_x=\lambda_5 u_t+\lambda_4 u_x.
\label{2e39} 
\end{equation}
We can eliminate $\lambda_4 u_x$ by the constant shift of $u\rightarrow u-\lambda_4/3$, which gives the KdV equation $\lambda_5 u_t-u_{xxx}+3 u u_x=0$. In the standard form of $\lambda_0=0$,  as the result of  the dual symmetry, by identifying  $\wp_{11}(u_1,u_2) \rightarrow u(x,t)$, $\dd u_1\rightarrow \dd x$ and $\dd u_2\rightarrow \dd t$, we have another KdV equation 
\begin{equation}
u_{xxx}-3 u u_x=\lambda_2 u_x+\lambda_1 u_t,
\label{2e40} 
\end{equation}
from Eq.(\ref{2e38}). 

We must notice that $u(x, t)=\wp_{xx}(x, t)=\partial_x^2 (-\log \sigma(x,t))$, expressed with the genus two hyperelliptic $\sigma$ function, is the solution but not the wave type solution, because $x$ and $t$ comes in the combination $X=x-v t \ (v: \text{const.})$ in the wave type solution.

In this way, we have the KdV equation and another KdV equation. As the Lie group structure of genus two hyperelliptic differential equations, we have sub structure of SO(2,1) and another SO(2,1) because each KdV equations have the SO(2,1) Lie group structure,

%%%%%%%%%%%%
\subsection{Differential equations of genus three hyperelliptic $\wp$ functions} 

We now move to the genus three case. The hyperelliptic curve in this case is given by 
\begin{equation}
C:\quad y_i^2=\sum_{k=0}^8 \lambda_k x_i^k , \quad (i=1, 2, 3).
\label{2e41}
\end{equation}
The Jacobi's inversion problem consists of solving the following system
\begin{equation}
\dd u_1=\sum_{i=1}^3 \frac{     \dd x_i}{y_i}, \qquad 
\dd u_2=\sum_{i=1}^3 \frac{x_i  \dd x_i}{y_i}, \qquad 
\dd u_3=\sum_{i=1}^3 \frac{x_i^2\dd x_i}{y_i}. 
\label{2e42}
\end{equation} 
Then we have 
\begin{equation}
\frac{\partial x_1}{\partial u_3}= \frac{         y_1}{(x_1-x_2)(x_1-x_3)},
\quad 
\frac{\partial x_1}{\partial u_2}=-\frac{(x_2+x_3)y_1}{(x_1-x_2)(x_1-x_3)} , 
\quad
\frac{\partial x_1}{\partial u_1}= \frac{ x_2 x_3 y_1}{(x_1-x_2)(x_1-x_3)} , 
\label{2e43}
\end{equation}
and $\{x_1, x_2, x_3\}, \{y_1, y_2, y_3\}$ cyclic permutation. In this case, 
\begin{align}
\dd(-\zeta_3)=
 &\sum_{i=1}^3 \frac{ \left(2 \lambda_8 x_i^4+\lambda_7 x_i^3\right)\dd x_i}{y_i}  ,
\label{2e44}\\
\dd(-\zeta_2)=
 &\sum_{i=1}^3 \frac{ \left(4 \lambda_8 x_i^5+3 \lambda_7 x_i^4+2 \lambda_6 x_i^3+\lambda_5 x_i^2\right)\dd x_i}{y_i}
\nonumber\\
 &-2\dd\left(\frac{y_1}{(x_1-x_2)(x_1-x_3)}
            +\frac{y_2}{(x_2-x_1)(x_2-x_3)}
            +\frac{y_3}{(x_3-x_1)(x_3-x_2)}\right)  ,
\label{2e45}\\
\dd(-\zeta_1)=
 &\sum_{i=1}^3 \frac{ \left(6\lambda_8 x_i^6+5\lambda_7 x_i^5+4\lambda_6 x_i^4+3\lambda_5 x_i^3
                           +2\lambda_4 x_i^2+ \lambda_3 x_i \right)\dd x_i}{y_i}
\nonumber\\
 &-2\dd\left(\frac{(x_1-x_2-x_3) y_1}{(x_1-x_2)(x_1-x_3)}
            +\frac{(x_2-x_3-x_1) y_2}{(x_2-x_1)(x_2-x_3)}
            +\frac{(x_3-x_1-x_2) y_3}{(x_3-x_1)(x_3-x_2)}\right)  .
\label{2e46}
\end{align}
For these $\zeta_{3}$, $\zeta_{2}$ and $\zeta_{1}$, we have checked the integrability condition
$\partial \zeta_i/\partial u_j=\partial \zeta_j/\partial u_i$, $(1\leq i < j \leq 3)$.
Just as the same as the genus two case, in order that differential equations become of the polynomial type, we must put
$\lambda_8=0$. In this case, we have 
\begin{equation}
\dd(-\zeta_3)=\lambda_7 \sum_{i=1}^3 \frac{ x_i^3 \dd x_i}{y_i} = \sum_{j=1}^3 \wp_{3j} \dd u_j ,
\label{2e47}
\end{equation}
which gives
\begin{align}
\widehat{\wp}_{33}
&=\frac{1}{\lambda_7} \wp_{33}=\frac{1}{\lambda_7} \frac{\partial (- \zeta_3)}{\partial u_3}
=x_1+x_2+x_3  , 
\label{2e48}\\
\widehat{\wp}_{32}
&=\frac{1}{\lambda_7} \wp_{32}=\frac{1}{\lambda_7} \frac{\partial (- \zeta_3)}{\partial u_2}
=-(x_1 x_2+x_2 x_3+x_3 x_1)  ,
\label{2e49}\\
\widehat{\wp}_{31}
&=\frac{1}{\lambda_7} \wp_{31}
=\frac{1}{\lambda_7} \frac{\partial (- \zeta_3)}{\partial u_1}=x_1 x_2 x_3  .
\label{2e50}
\end{align}
Then we have the following differential equations\cite{Baker4,Buchstaber1,Buchstaber2}
\begin{align}
 1)\quad & \wp_{3333}-\frac{3}{2}\wp_{33}^2
  =\lambda_7 \wp_{32}+\lambda_6\wp_{33}+\frac{1}{2} \lambda_7 \lambda_5 ,
\label{2e51}\\
 2)\quad & \wp_{3332}-\frac{3}{2}\wp_{33} \wp_{32}
  =\frac{3}{2}\lambda_7 \wp_{31} -\frac{1}{2}\lambda_7 \wp_{22}+\lambda_6\wp_{32} ,
\label{2e52}\\
 3)\quad & \wp_{3331}-\frac{3}{2}\wp_{33}\wp_{31}
  =-\frac{1}{2}\lambda_7 \wp_{21} + \lambda_6\wp_{31} , 
\label{2e53}\\
 4)\quad & \wp_{3322}-\frac{1}{2}\wp_{33} \wp_{22}-\wp_{32}^2
  =-\frac{1}{2}\lambda_7\wp_{21} +\lambda_6 \wp_{31}+\frac{1}{2}\lambda_5\wp_{32}  ,
\label{2e54}\\
 5)\quad & \wp_{3321}-\frac{1}{2}\wp_{33} \wp_{21}-\wp_{32}\wp_{31}
  =\frac{1}{2}\lambda_5\wp_{31}  , 
\label{2e55}\\
 6)\quad & \wp_{3311}-\frac{1}{2}\wp_{33}\wp_{11}-\wp_{31}^2 
  =\frac{1}{2}\Delta  ,
\label{2e56} \\
 7)\quad & \wp_{3222}-\frac{3}{2}\wp_{32} \wp_{22}
  =-\frac{3}{2}\lambda_7\wp_{11} 
+\lambda_5 \wp_{31}+\lambda_4\wp_{32}-\frac{1}{2}\lambda_3\wp_{33} -\lambda_7 \lambda_2 , 
\label{2e57}\\
 8)\quad & \wp_{3221}-\frac{1}{2}\wp_{31} \wp_{22}-\wp_{32}\wp_{21}
  =-\frac{1}{2}\Delta +\lambda_4\wp_{31} -\frac{1}{2}\lambda_7\lambda_1 ,
\label{2e58}\\
 9)\quad & \wp_{3211}-\frac{1}{2}\wp_{32} \wp_{11}-\wp_{31}\wp_{21}
  =\frac{1}{2}\lambda_3\wp_{31} -\lambda_7\lambda_0 ,
\label{2e59}\\
10)\quad & \wp_{3111}-\frac{3}{2}\wp_{31}\wp_{11}
  =\lambda_2 \wp_{31} 
-\frac{1}{2}\lambda_1 \wp_{32} +\lambda_0 \wp_{33} ,
\label{2e60}\\
11)\quad & \wp_{2222}-\frac{3}{2}\wp_{22}^2
  =3 \Delta
-3\lambda_6\wp_{11}+\lambda_5 \wp_{21}+\lambda_4\wp_{22}+\lambda_3\wp_{32} 
-3\lambda_2\wp_{33}\nonumber\\
&
-2\lambda_6 \lambda_2+\frac{1}{2}\lambda_5 \lambda_3
-\frac{3}{2}\lambda_7\lambda_1 ,
\label{2e61}\\
12)\quad & \wp_{2221}-\frac{3}{2}\wp_{22} \wp_{21}
  =-\frac{1}{2}\lambda_5 \wp_{11}
+\lambda_4\wp_{21} +\lambda_3 \wp_{31} -\frac{3}{2}\lambda_1\wp_{33} 
-2 \lambda_7 \lambda_0-\lambda_6 \lambda_1 ,
\label{2e62}\\
13)\quad & \wp_{2211}-\frac{1}{2}\wp_{22}\wp_{11}-\wp_{21}^2
  =\frac{1}{2}\lambda_3 \wp_{21}+\lambda_2 \wp_{31} -\frac{1}{2}\lambda_1 \wp_{32}-2\lambda_0 \wp_{33}
-2 \lambda_6 \lambda_0 ,
\label{2e63}\\
14)\quad & \wp_{2111}-\frac{3}{2}\wp_{21}\wp_{11}
  =\lambda_2\wp_{21}+\frac{3}{2}\lambda_1 \wp_{31}
-\frac{1}{2}\lambda_1 \wp_{22}-2\lambda_0 \wp_{32}-\lambda_5 \lambda_0 ,
\label{2e64}\\
15)\quad & \wp_{1111}-\frac{3}{2}\wp_{11}^2
  =\lambda_2 \wp_{11}
+\lambda_1\wp_{21} +4\lambda_0 \wp_{31}  -3\lambda_0 \wp_{22} 
-2\lambda_4 \lambda_0+\frac{1}{2}\lambda_3 \lambda_1 ,
\label{2e65}
\end{align}
where $\Delta=\wp_{32}\wp_{21}-\wp_{31}\wp_{22}-\wp_{33}\wp_{11}+\wp_{31}^2$.

Just as in genus two case, if we take $\lambda_0=0$ as the standard form 
of the hyperelliptic curve, the set of differential equations have some dual symmetry 
Eq.({\ref{2e51}) $\leftrightarrow$ Eq.({\ref{2e65}),
Eq.({\ref{2e52}) $\leftrightarrow$ Eq.({\ref{2e64}), 
Eq.({\ref{2e53}) $\leftrightarrow$ Eq.({\ref{2e60}),  etc., 
under $u_3 \leftrightarrow \pm u_1$, $u_2 \leftrightarrow \pm u_2$, 
$\lambda_1  \leftrightarrow \lambda_7$, $\lambda_2  \leftrightarrow \lambda_6$, 
$\lambda_3  \leftrightarrow \lambda_5$ and $\lambda_4  \leftrightarrow \lambda_4$.
In this standard form of $\lambda_0=0$, Eq.(\ref{2e51}) and Eq.(\ref{2e65}) become KdV equation Eq.(\ref{2e39}) with $\lambda_j \rightarrow \lambda_{j+2}$ and another KdV equation Eq.(\ref{2e40}), respectively.

While if we take $\lambda_1=0$ as the standard form, by identifying $\wp_{11} \rightarrow u$, $\dd u_1 \rightarrow \dd x$, $\dd u_2 \rightarrow \dd y$ and $\dd u_3 \rightarrow \dd t$, by differentiating Eq.(\ref{2e65}) with $u_1$ twice, we have KP equation  
\begin{equation}
\big(u_{xxx}-3u u_x-\lambda_2 u_x-4 \lambda_0 u_t\big)_x=-3 \lambda_0 u_{yy}  .
\label{2e66}
\end{equation}
In this way, Eq.({\ref{2e65}) becomes the KdV equation in the $\lambda_0=0$ standard form, and the same Eq.({\ref{2e65}) becomes the KP equation in the $\lambda_1=0$ standard form. Then the difference of the KdV equation and the KP equation comes from the choice of standard form of the hyperelliptic curve. Therefore, the KdV equation and the KP equation belong to the same family in this approach. 

By differentiating Eq.(\ref{2e61}) with $u_2$ twice, we have the following three variables differential equation  
\begin{equation}
\big(u_{xxx}-3 u u_x -\lambda_4 u_{x}-\lambda_5 u_t \big)_x=3 \Delta_{xx}
-3 \lambda_6 u_{tt}+\lambda_3 u_{xy} -3 \lambda_2 u_{yy} ,
\label{2e67}
\end{equation}
by identifying $\wp_{22}\rightarrow u$, $\dd u_1\rightarrow \dd t$, $\dd u_2 \rightarrow \dd x$ and $\dd u_3\rightarrow \dd y$. If we consider the special hyperelliptic curve with $\lambda_6=0$ and $\lambda_3=0$, Eq.(\ref{2e66}) becomes the KP equation except $\Delta_{xx}$ term in the form 
\begin{equation}
\big(u_{xxx}-3 u u_x -\lambda_4 u_{x}-\lambda_5 u_t\big)_x+3 \lambda_2 u_{yy} =3 \Delta_{xx} , 
\label{2e68}
\end{equation}
and we have checked that $\Delta_{xx} \ne 0$ even for this special hyperelliptic curve. Then we have three variables new type integrable differential equation, which is KP type but is different from KP equation itself. 

%%%%%%%%%%%%%%%%%%%%%%%%%%%%%%%%%%%%%%%%%%%%%%%%%%%%%%%%%%%
\section{Differential Equations of Genus Four Hyperelliptic $\wp$ Functions} 
\setcounter{equation}{0}

Now let us consider the genus four case. The hyperelliptic curve in this case is given by 
\begin{equation}
C:\quad y_i^2=\sum_{k=0}^{10} \lambda_{k} x_i^{k} , \quad (i=1, 2, 3, 4).
\label{3e1}
\end{equation}
The Jacobi's inversion problem consists of solving the following system
\begin{eqnarray}
\dd u_1=\sum_{i=1}^4 \frac{      \dd x_i}{y_i}, \qquad 
\dd u_2=\sum_{i=1}^4 \frac{x_i   \dd x_i}{y_i}, \qquad 
\dd u_3=\sum_{i=1}^4 \frac{x_i^2 \dd x_i}{y_i}, \qquad 
\dd u_4=\sum_{i=1}^4 \frac{x_i^3 \dd x_i}{y_i}.
\label{3e2}
\end{eqnarray} 
Then we have 
\begin{align}
\frac{\partial x_1}{\partial u_4}&= \frac{y_1}{(x_1-x_2)(x_1-x_3)(x_1-x_4)} , &  
\frac{\partial x_1}{\partial u_3}&=-\frac{(x_2+x_3+x_4) y_1}{(x_1-x_2)(x_1-x_3)(x_1-x_4)} , 
\nonumber\\
\frac{\partial x_1}{\partial u_2}&= \frac{(x_2 x_3+x_3 x_4+x_4 x_2) y_1}
{(x_1-x_2)(x_1-x_3)(x_1-x_4)} ,  & 
\frac{\partial x_1}{\partial u_1}&=-\frac{x_2 x_3 x_4 y_1}{(x_1-x_2)(x_1-x_3)(x_1-x_4)}  , 
\label{3e3}
\end{align}
and $\{x_1,x_2,x_3,x_4\}, \{y_1,y_2,y_3,y_4\}$ cyclic permutation. In this case,
\begin{align}
\dd(-\zeta_4)=
 &\sum_{i=1}^4
\frac{ \left(2 \lambda_{10} x_i^5+\lambda_9 x_i^4\right)\dd x_i}{y_i}  ,
\label{3e4}\\
\dd(-\zeta_3)=
 &\sum_{i=1}^4
\frac{ \left(4 \lambda_{10} x_i^6+3 \lambda_9 x_i^5+2 \lambda_8 x_i^4+\lambda_7 x_i^3\right)\dd x_i}
{y_i}\nonumber\\
 &    -2\dd\bigg(\frac{y_1}{(x_1-x_2)(x_1-x_3)(x_1-x_4)}+\frac{y_2}{(x_2-x_1)(x_2-x_3)(x_2-x_4)} \nonumber\\
 &\hspace*{10mm}+\frac{y_3}{(x_3-x_1)(x_3-x_2)(x_3-x_4)}
                +\frac{y_4}{(x_4-x_1)(x_4-x_2)(x_4-x_3)}\bigg)  ,
\label{3e5}\\
\dd(-\zeta_2)=
 &\sum_{i=1}^4
\frac{ \left(6 \lambda_{10} x_i^7+5 \lambda_9 x_i^6+4 \lambda_8 x_i^5+3 \lambda_7 x_i^4
            +2 \lambda_6    x_i^3+  \lambda_5 x_i^2  \right)\dd x_i}{y_i}\nonumber\\
 &    -2\dd\bigg(\frac{(x_1-x_2-x_3-x_4) y_1}{(x_1-x_2)(x_1-x_3)(x_1-x_4)}
                +\frac{(x_2-x_3-x_4-x_1) y_2}{(x_2-x_1)(x_2-x_3)(x_2-x_4)}\nonumber\\ 
 &\hspace*{10mm}+\frac{(x_3-x_4-x_1-x_2) y_3}{(x_3-x_1)(x_3-x_2)(x_3-x_4)}
                +\frac{(x_4-x_1-x_2-x_3) y_4}{(x_4-x_1)(x_4-x_2)(x_4-x_3)}\bigg)  ,
\label{3e6}\\
\dd(-\zeta_1)=
 &\sum_{i=1}^4
\frac{ \left(8 \lambda_{10} x_i^8+7 \lambda_{9} x_i^7+6 \lambda_8 x_i^6+5 \lambda_7 x_i^5
            +4 \lambda_6    x_i^4+3 \lambda_5   x_i^3+2 \lambda_4 x_i^2+  \lambda_3 x_i\right)\dd x_i}{y_i}\nonumber\\
 &-2\dd\left(
                 \frac{\left(x_1^2-x_1(x_2+x_3+x_4)+(x_2x_3+x_3 x_4+x_4x_2)\right)y_1}{(x_1-x_2)(x_1-x_3)(x_1-x_4)}\right.\nonumber\\
 &\hspace*{10mm}+\frac{\left(x_2^2-x_2(x_3+x_4+x_1)+(x_3x_4+x_4 x_1+x_1x_3)\right)y_2}{(x_2-x_1)(x_2-x_3)(x_2-x_4)}\nonumber\\
 &\hspace*{10mm}+\frac{\left(x_3^2-x_3(x_4+x_1+x_2)+(x_4x_1+x_1 x_2+x_2x_4)\right)y_3}{(x_3-x_1)(x_3-x_2)(x_3-x_4)}\nonumber\\
 &\hspace*{10mm}\left.+\frac{\left(x_4^2-x_4(x_1+x_2+x_3)+(x_1x_2+x_2 x_3+x_3x_1)\right) y_4}{(x_4-x_1)(x_4-x_2)(x_4-x_3)}\right).
\label{3e7}
\end{align}
For these $\zeta_{4}$, $\zeta_{3}$, $\zeta_{2}$ and $\zeta_{1}$, we have checked the integrability condition $\partial \zeta_i/\partial u_j=\partial \zeta_j/\partial u_i$, $(1\leq i < j \leq 4)$. Just as the same as the genus two and genus three cases, in order that differential equations become of the polynomial type, we must take $\lambda_{10}=0$. In this case, we have 
\begin{equation}
\dd(-\zeta_4)=\lambda_9 \sum_{i=1}^4 \frac{ x_i^4\dd x_i}{y_i}
=\sum_{j=1}^4 \wp_{4j}\dd u_j  ,
\label{3e8}
\end{equation}
which gives
\begin{align}
\widehat{\wp}_{44}
 &=\frac{1}{\lambda_9}\wp_{44}
  =\frac{1}{\lambda_9}\frac{\partial (- \zeta_4)}{\partial u_4}
  =x_1+x_2+x_3+x_4   ,
\label{3e9}\\
\widehat{\wp}_{43}
 &=\frac{1}{\lambda_9}\wp_{43}
  =\frac{1}{\lambda_9}\frac{\partial (- \zeta_4)}{\partial u_3}
  =-(x_1 x_2+x_1 x_3+x_1 x_4+x_2 x_3+x_2 x_4+x_3 x_4)   ,
\label{3e10}\\
\widehat{\wp}_{42}
 &=\frac{1}{\lambda_9}\wp_{42}
  =\frac{1}{\lambda_9}\frac{\partial (- \zeta_4)}{\partial u_2}
  =x_1 x_2 x_3+x_1 x_2 x_4+x_1 x_3 x_4+x_2 x_3 x_4  ,
\label{3e11}\\
\widehat{\wp}_{41}
 &=\frac{1}{\lambda_9}\wp_{41}
  =\frac{1}{\lambda_9}\frac{\partial (- \zeta_4)}{\partial u_1}
  =-x_1 x_2 x_3 x_4   .
\label{3e12}
\end{align}
Then we have the following differential equations
\begin{align}
 1)\quad & \wp_{4444}-\frac{3}{2}\wp_{44}^2
  = \lambda_9 \wp_{43}+\lambda_8\wp_{44}+\frac{1}{2}\lambda_9 \lambda_7   ,
\label{3e13}\\
 2)\quad & \wp_{4443}-\frac{3}{2}\wp_{44}\wp_{43}
  = \frac{3}{2}\lambda_9 \wp_{42}-\frac{1}{2}\lambda_9 \wp_{33}+\lambda_8\wp_{43}  ,
\label{3e14}\\
 3)\quad & \wp_{4442}-\frac{3}{2}\wp_{44}\wp_{42}
  = \frac{3}{2}\lambda_9 \wp_{41}-\frac{1}{2}\lambda_9 \wp_{32} 
+\lambda_8 \wp_{42}  ,
\label{3e15}\\
 4)\quad & \wp_{4441}-\frac{3}{2}\wp_{44}\wp_{41}
  =-\frac{1}{2}\lambda_9\wp_{31} +\lambda_8 \wp_{41}   ,
\label{3e16}\\
 5)\quad & \wp_{4433}-\frac{1}{2}\wp_{44} \wp_{33}-\wp_{43}^2
  = \frac{3}{2}\lambda_9\wp_{41}-\frac{1}{2}\lambda_9\wp_{32}+\lambda_8\wp_{42}+\frac{1}{2}\lambda_7\wp_{43}   ,
\label{3e17}\\
 6)\quad & \wp_{4432}-\frac{1}{2}\wp_{44}\wp_{32}-\wp_{43}\wp_{42} 
  =-\frac{1}{2}\lambda_9 \wp_{31}+\lambda_8 \wp_{41}+\frac{1}{2}\lambda_7 \wp_{42}  ,
\label{3e18}\\
 7)\quad  & \wp_{4431}-\frac{1}{2}\wp_{44} \wp_{31}-\wp_{43} \wp_{41}
  = \frac{1}{2}\lambda_7\wp_{41}   ,
\label{3e19}\\
 8)\quad & \wp_{4422}-\frac{3}{2}\wp_{42}^2
  = \frac{1}{2} \Delta_1+\frac{1}{2}\lambda_7 \wp_{41}   ,
\label{3e20}\\
 9)\quad & \wp_{4421}-\frac{3}{2}\wp_{42} \wp_{41}
  = \frac{1}{2} \Delta_8  ,
\label{3e21}\\
10)\quad & \wp_{4411}-\frac{3}{2}\wp_{41}^2
  = \frac{1}{2} \Delta_9  ,
\label{3e22}\\
11)\quad & \wp_{4333}-\frac{3}{2}\wp_{43}\wp_{33}
  = \frac{3}{2}\lambda_9\wp_{31}-\frac{3}{2}\lambda_9\wp_{22}
+\lambda_8\wp_{41}+\lambda_7 \wp_{42}+\lambda_6\wp_{43}\nonumber\\
&-\frac{1}{2}\lambda_5\wp_{44}-\lambda_9\lambda_4  ,
\label{3e23}\\
12)\quad & \wp_{4332}-\frac{1}{2}\wp_{42} \wp_{33}-\wp_{43} \wp_{32}
  =-\frac{1}{2} \Delta_2 -\lambda_9\wp_{21} +\lambda_7 \wp_{41}+\lambda_6\wp_{42}
-\frac{1}{2} \lambda_9 \lambda_3   ,
\label{3e24}\\
13)\quad & \wp_{4331}-\frac{1}{2}\wp_{41}\wp_{33}-\wp_{43}\wp_{31}
  = \frac{1}{2} \Delta_3+\frac{1}{2}\lambda_9 \wp_{11}+\lambda_6 \wp_{41}   ,
\label{3e25}\\
14)\quad & \wp_{4322}-\frac{1}{2}\wp_{43}\wp_{22}-\wp_{42}\wp_{32}
  = \frac{1}{2} \Delta_3-\lambda_9\wp_{11}+\lambda_6 \wp_{41}
+\frac{1}{2}\lambda_5 \wp_{42}-\lambda_9 \lambda_2   ,
\label{3e26}\\
15)\quad & \wp_{4321}-\frac{1}{2}\wp_{43}\wp_{21}-\frac{1}{2}\wp_{42}\wp_{31}
-\frac{1}{2}\wp_{41}\wp_{32}
  = \frac{1}{2}\lambda_5\wp_{41} -\frac{1}{2}\lambda_9\lambda_1   ,
\label{3e27}\\
16)\quad & \wp_{4311}-\frac{3}{2}\wp_{41} \wp_{31}
  = \frac{1}{2}\Delta_{10}
-\lambda_9\lambda_0   ,
\label{3e28}\\
17)\quad & \wp_{4222}-\frac{3}{2}\wp_{42}\wp_{22}
  = \frac{3}{2} \Delta_4+\lambda_5 \wp_{41} 
+\lambda_4 \wp_{42} -\frac{1}{2}\lambda_3 \wp_{43}+\lambda_2 \wp_{44}
-\lambda_9\lambda_1    ,
\label{3e29}\\
18)\quad & \wp_{4221}-\frac{1}{2}\wp_{41} \wp_{22}-\wp_{42} \wp_{21}
  = \frac{1}{2} \Delta_5+\lambda_4\wp_{41} +\frac{1}{2}\lambda_1 \wp_{44} -\lambda_9\lambda_0 , 
\label{3e30}\\
19)\quad & \wp_{4211}-\frac{1}{2}\wp_{42}\wp_{11}-\wp_{41}\wp_{21}
 =\frac{1}{2}\lambda_3 \wp_{41}+\lambda_0 \wp_{44}  , 
\label{3e31}\\
20)\quad & \wp_{4111}-\frac{3}{2}\wp_{41} \wp_{11}
  = \lambda_2 \wp_{41} -\frac{1}{2}\lambda_1 \wp_{42} +\lambda_0 \wp_{43} , 
\label{3e32}\\
21)\quad & \wp_{3333}-\frac{3}{2}\wp_{33}^2
  =3 \Delta_2-3\lambda_9 \wp_{21}+4\lambda_8 \wp_{31}
-3\lambda_8 \wp_{22}+\lambda_7 \wp_{32}
+\lambda_6 \wp_{33}+\lambda_5 \wp_{43}\nonumber\\
&-3\lambda_4 \wp_{44}+\frac{1}{2}\lambda_7 \lambda_5
-2\lambda_8 \lambda_4-\frac{3}{2} \lambda_9\lambda_3  ,
\label{3e33}\\
22)\quad & \wp_{3332}-\frac{3}{2}\wp_{33}\wp_{32}
  =-\frac{3}{2} \Delta_3-\frac{3}{2}\lambda_9 \wp_{11}-2\lambda_8 \wp_{21}
+\frac{3}{2}\lambda_7 \wp_{31}
-\frac{1}{2}\lambda_7 \wp_{22}+\lambda_6 \wp_{32} \nonumber\\
&+\lambda_5 \wp_{42} -\frac{3}{2}\lambda_3 \wp_{44}
-\lambda_8 \lambda_3-2\lambda_9\lambda_2 ,
\label{3e34}\\
23)\quad & \wp_{3331}-\frac{3}{2}\wp_{33}\wp_{31}
  = \frac{3}{2} \Delta_4+\lambda_8 \wp_{11}-\frac{1}{2}\lambda_7 \wp_{21}
+\lambda_6 \wp_{31}+\lambda_5 \wp_{41}-\lambda_9\lambda_1 ,
\label{3e35}\\
24)\quad & \wp_{3322}-\frac{1}{2}\wp_{33}\wp_{22}-\wp_{32}^2
  =-\frac{3}{2} \Delta_4-2\lambda_8 \wp_{11}-\frac{1}{2}\lambda_7 \wp_{21}
+\lambda_6 \wp_{31}\nonumber\\
&+\frac{1}{2}\lambda_5 \wp_{41}+\frac{1}{2}\lambda_5 \wp_{32}
+\lambda_4 \wp_{42}-\frac{1}{2}\lambda_3 \wp_{43}-2\lambda_2 \wp_{44}
-2\lambda_8 \lambda_2-2\lambda_9\lambda_1  ,
\label{3e36}\\
25)\quad & \wp_{3321}-\frac{1}{2}\wp_{33}\wp_{21}-\wp_{32}\wp_{31}
  = \frac{1}{2} \Delta_5 
+\frac{1}{2}\lambda_5 \wp_{31}+\lambda_4 \wp_{41}-\lambda_1 \wp_{44}
\nonumber\\
&-\lambda_8 \lambda_1-2\lambda_9\lambda_0   , 
\label{3e37}\\
26)\quad & \wp_{3311}-\frac{3}{2}\wp_{31}^2
  = \frac{1}{2} \Delta_6+\frac{1}{2}\lambda_3 \wp_{41}-2\lambda_0 \wp_{44}
-2\lambda_8 \lambda_0  ,
\label{3e38}\\
27)\quad & \wp_{3222}-\frac{3}{2}\wp_{32}\wp_{22}
  =-\frac{3}{2} \Delta_5-\frac{3}{2}\lambda_7 \wp_{11} +\lambda_5 \wp_{31}  
+\lambda_4 \wp_{32}+\frac{3}{2}\lambda_3 \wp_{42}\nonumber\\
&-\frac{1}{2}\lambda_3 \wp_{33}-2\lambda_2 \wp_{43}
-\frac{3}{2}\lambda_1 \wp_{44}
-2\lambda_8 \lambda_1-\lambda_7 \lambda_2-3\lambda_9\lambda_0   , 
\label{3e39}\\
28)\quad & \wp_{3221}-\frac{1}{2}\wp_{31}\wp_{22}-\wp_{32}\wp_{21}
  =-\frac{1}{2} \Delta_7+\lambda_4 \wp_{31}
+\lambda_3 \wp_{41}-\lambda_1 \wp_{43}-\lambda_0 \wp_{44}\nonumber\\
&-2\lambda_8 \lambda_0-\frac{1}{2}\lambda_7 \lambda_1 ,
\label{3e40}\\
29)\quad & \wp_{3211}-\frac{1}{2}\wp_{32}\wp_{11}-\wp_{31}\wp_{21}
  = \frac{1}{2}\lambda_3 \wp_{31}+\lambda_2 \wp_{41}-\frac{1}{2}\lambda_1 \wp_{42}
-\lambda_0 \wp_{43} -\lambda_7 \lambda_0  , 
\label{3e41}\\
30)\quad & \wp_{3111}-\frac{3}{2}\wp_{31}\wp_{11}
  = \lambda_2 \wp_{31}+\frac{3}{2}\lambda_1 \wp_{41}-\frac{1}{2}\lambda_1 \wp_{32}-3\lambda_0 \wp_{42}
+\lambda_0 \wp_{33} , 
\label{3e42}\\
31)\quad & \wp_{2222}-\frac{3}{2}\wp_{22}^2
  =3\Delta_7-3\lambda_6 \wp_{11}+\lambda_5 \wp_{21}
+\lambda_4 \wp_{22}+\lambda_3 \wp_{32}+4\lambda_2 \wp_{42}\nonumber\\
&-3\lambda_2 \wp_{33}-3\lambda_1 \wp_{43}-3\lambda_0 \wp_{44}
-4\lambda_8 \lambda_0-\frac{3}{2}\lambda_7 \lambda_1
-2\lambda_6 \lambda_2+\frac{1}{2}\lambda_5 \lambda_3 , 
\label{3e43}\\
32)\quad & \wp_{2221}-\frac{3}{2}\wp_{22}\wp_{21}
  =-\frac{1}{2}\lambda_5 \wp_{11}+\lambda_4 \wp_{21}+\lambda_3 \wp_{31}+\lambda_2 \wp_{41}
+\frac{3}{2}\lambda_1 \wp_{42}-\frac{3}{2}\lambda_1 \wp_{33}\nonumber\\
&-3\lambda_0 \wp_{43}-2\lambda_7 \lambda_0-\lambda_6 \lambda_1   ,
\label{3e44}\\
33)\quad & \wp_{2211}-\frac{1}{2}\wp_{22}\wp_{11}-\wp_{21}^2
  = \frac{1}{2}\lambda_3 \wp_{21}+\lambda_2 \wp_{31}
+\frac{3}{2}\lambda_1 \wp_{41}-\frac{1}{2}\lambda_1 \wp_{32}+\lambda_0 \wp_{42}\nonumber\\
&-2\lambda_0 \wp_{33}-2\lambda_6 \lambda_0  ,
\label{3e45}
\\
34)\quad & \wp_{2111}-\frac{3}{2}\wp_{21}\wp_{11}
  = \lambda_2 \wp_{21}+\frac{3}{2}\lambda_1 \wp_{31}-\frac{1}{2}\lambda_1 \wp_{22}+3\lambda_0 \wp_{41}
-2\lambda_0 \wp_{32}
-\lambda_5 \lambda_0  ,
\label{3e46}\\
35)\quad & \wp_{1111}-\frac{3}{2}\wp_{11}^2
  = \lambda_2 \wp_{11}+\lambda_1 \wp_{21}+4\lambda_0 \wp_{31}-3\lambda_0 \wp_{22}
-2\lambda_4 \lambda_0+\frac{1}{2}\lambda_3 \lambda_1 ,
\label{3e47}
\end{align}
where 
\begin{align*}
\Delta_1   &=\wp_{44}\wp_{31}-\wp_{43}\wp_{41}+\wp_{43}\wp_{32}-\wp_{42}\wp_{33}, 
&
\Delta_2   &=\Delta_1-\wp_{44}\wp_{22}+\wp_{42}^2,  
\\
\Delta_3   &=\wp_{44}\wp_{21}-\wp_{42}\wp_{41}-\wp_{43}\wp_{31}+\wp_{41}\wp_{33},  
&
\Delta_4   &=\wp_{44}\wp_{11}-\wp_{41}^2      -\wp_{42}\wp_{31}+\wp_{41}\wp_{32}, 
\\
\Delta_5   &=\wp_{43}\wp_{11}-\wp_{41}\wp_{31}-\wp_{42}\wp_{21}+\wp_{41}\wp_{22},  
& 
\Delta_6   &=\wp_{42}\wp_{11}-\wp_{41}\wp_{21}+\wp_{32}\wp_{21}-\wp_{31}\wp_{22} , 
\\ 
\Delta_7   &=\Delta_6-\wp_{33}\wp_{11}+\wp_{31}^2, 
& 
\Delta_8   &=\wp_{43}\wp_{31}-\wp_{41}\wp_{33}, 
\\
\Delta_9   &=\wp_{42}\wp_{31}-\wp_{41}\wp_{32}, 
&  
\Delta_{10}&=\wp_{42}\wp_{21}-\wp_{41}\wp_{22} . 
\end{align*}
These $\Delta_i$ have the symmetry
$\Delta_1 \leftrightarrow \Delta_6$, 
$\Delta_2 \leftrightarrow \Delta_7$,
$\Delta_3 \leftrightarrow \Delta_5$, 
$\Delta_4 \leftrightarrow \Delta_4$,
$\Delta_8 \leftrightarrow \Delta_{10}$ and
$\Delta_9 \leftrightarrow \Delta_9$, under
$\dd u_1 \leftrightarrow \pm \dd u_4$ and
$\dd u_2 \leftrightarrow \pm \dd u_3$.

Just as in genus two and three cases, in the standard form of the hyperelliptic curve of $\lambda_0=0$, the set of differential equations have the dual symmetry 
Eq.({\ref{3e13}) $\leftrightarrow$ Eq.({\ref{3e47}),
Eq.({\ref{3e14}) $\leftrightarrow$ Eq.({\ref{3e46}), 
Eq.({\ref{3e15}) $\leftrightarrow$ Eq.({\ref{3e42}),  etc., 
under $u_4 \leftrightarrow \pm u_1$, $u_3 \leftrightarrow \pm u_2$, 
$\lambda_1  \leftrightarrow \lambda_9$, $\lambda_2  \leftrightarrow \lambda_8$, 
$\lambda_3  \leftrightarrow \lambda_7$, $\lambda_4  \leftrightarrow \lambda_6$ and
$\lambda_5  \leftrightarrow \lambda_5$.

In the standard form of $\lambda_0=0$, the differential equation of Eq.(\ref{3e13}) and Eq.(\ref{3e47}) are KdV equation Eq.(\ref{2e39}) with $\lambda_j \rightarrow \lambda_{j+4}$ and another KdV equation Eq.(\ref{2e40}),respectively. While in the standard form of $\lambda_1=0$, the differential equation Eq.(\ref{3e47}) is KP equation Eq.(\ref{2e66}).

By differentiating Eq.(\ref{3e33}) with $u_3$ twice, we have four variables differential equation, which is KP type equation except the term $(\Delta_2)_{xx}(\ne 0)$ in the form 
\begin{equation}
\big(u_{xxx}-3 u u_x -\lambda_7 u_t-\lambda_6 u_x \big)_x 
= 3(\Delta_2)_{xx} - 3\lambda_9u_{zt} + 4\lambda_8u_{zx} - 3\lambda_8u_{tt} + \lambda_5u_{xy} - 3\lambda_4u_{yy},
\label{3e48}
\end{equation}
by identifying
$\wp_{33}\rightarrow u$,
$\dd u_1\rightarrow \dd z$, 
$\dd u_2\rightarrow \dd t$, 
$\dd u_3\rightarrow \dd x$ and
$\dd u_4\rightarrow \dd y$. 
Then we have four variables KP type new integrable differential equation. Eq.(\ref{3e43}) is four variables another KP type differential equation.

%%%%%%%%%%%%%%%%%%%%%%%%%%%%%%%%%%%%%%%%%%%%%%%%%%%%%%%%%%%
\section{Properties of Hyperelliptic Differential Equations} 
\setcounter{equation}{0}

\subsection{Some dual symmetry for the set of differential equations}

In the previous sections, we have explained the symmetry of differential equations, that is, in the standard form of $\lambda_{2g+2}=0$ and $\lambda_{0}=0$ in the hyperelliptic curve, the set of differential equations have some dual symmetry under 
\begin{equation}
\wp_{jk}   \leftrightarrow \wp_{g+1-j,g+1-k},            \quad 
\wp_{jklm} \leftrightarrow \wp_{g+1-j,g+1-k,g+1-l,g+1-m},\quad   
\lambda_{k}\leftrightarrow \tilde{\lambda}_k=\lambda_{2g+2-k}. 
\label{dual_symm_00}
\end{equation}

The standard form of the hyperelliptic curve is given by 
\begin{equation}
C: \quad y_i^2 = \lambda_{2g+1}x_i^{2g+1} + \lambda_{2g}x_i^{2g} + \cdots + \lambda_{2}x_i^{2} + \lambda_{1}x_i.
\label{4e1}
\end{equation}
If we change variables in the form 
$\tilde{x}_i=\dfrac{1}{x_i}$,\
$\tilde{y}_i=\dfrac{y_i}{x_i^{g+1}}$, \
$\tilde{\lambda}_{k}=\lambda_{2g+2-k}$, we can rewrite the curve in the form  
\begin{equation}
\tilde{C}: \quad  \tilde{y}_i^2=\tilde{\lambda}_{2g+1} \tilde{x}_i^{2g+1}
+\tilde{\lambda}_{2g}\tilde{x}_i^{2g}+\cdots 
+\tilde{\lambda}_{2} \tilde{x}_i^{2} +\tilde{\lambda}_{1}\tilde{x}_i  .
\label{4e2}
\end{equation}
Then we have
\begin{equation}
\dd\tilde{u}_j=\sum_{i=1}^g \frac{ \tilde{x}_{i}^{j-1}\dd \tilde{x}_i } { \tilde{y}_i }
=-\sum_{i=1}^g \frac{ x_i^{g-j}\dd x_i }{ y_i}
=-\dd u_{g+1-j}  ,
\label{4e3}
\end{equation}
that is, $\dd\tilde{u}_g=-\dd u_{1}$, \ $\dd\tilde{u}_{g-1}=-\dd u_{2}$, \ $\cdots$, \ 
$\dd\tilde{u}_2=-\dd u_{g-1}$, \ and \  
$\dd\tilde{u}_1=-\dd u_{g}$. 

From the curve Eq.(\ref{4e2}), we construct hyperelliptic sigma function $\tilde{\sigma}$. While we construct $\sigma$ from the curve Eq.(\ref{4e1}). But the difference between Eq.(\ref{4e1}) and Eq.(\ref{4e2}) is only the choice of the local variable, so that $\sigma$ function and $\tilde{\sigma}$ function is essentially the same, then we have  
$\dfrac{\partial(-\log \tilde{\sigma})}{\partial \tilde{u}_j}
=\dfrac{\partial(-\log \sigma)        }{\partial \tilde{u}_j}
=(-\zeta_{\tilde{j}})$. 
Then $\dd u_j \leftrightarrow -\dd \tilde{u}_j$ is equivalent to
$\wp_{jk}  \leftrightarrow (-1)^2 \wp_{\tilde{\mathstrut j}\tilde{\mathstrut k}}=\wp_{\tilde{\mathstrut j}\tilde{\mathstrut k}}$ and 
$\wp_{jklm}\leftrightarrow (-1)^4 \wp_{\tilde{\mathstrut j}\tilde{\mathstrut k}\tilde{\mathstrut l}\tilde{\mathstrut m}}=\wp_{\tilde{\mathstrut j}\tilde{\mathstrut k}\tilde{\mathstrut l}\tilde{\mathstrut m}}$.
Therefore, we conclude that the set of differential equations have some dual symmetry under (\ref{dual_symm_00}). 

%%%%%%%%%%%%%%%%%%%%%%%%%%%%%%%%%%%%%%%%%%%%%%%%%%%%%%%%%%%%%
%%%%%%%%%%%%%%%%%%%%%%%%%%%%%%%%%%%%%%%%%%%%%%%%%%%%%%%%%%%
\subsection{Some differential equations for general genus } 
Buchstaber et al. have shown the quite interesting result that one family of differential equations always exist for general genus~\cite {Buchstaber1,Buchstaber2}, and it is really the family of KdV equation in 1997~\cite{Buchstaber1}. We sketch the proof of their result.  We start from
\begin{align}
\dd(-\zeta_{g-1})
 &=\sum_{i=1}^g
     \frac{\left(\lambda_{2g-1}x_i^{g-1}+2\lambda_{2g}x_i^{g}+3\lambda_{2g+1}x_i^{g+1}\right) \dd x_i}{y_i} 
     -2 \dd \left(\sum_{i=1}^g \frac{y_i}{F'(x_i)}\right)
\notag\\
 &=\sum_{i=1}^g
     \frac{\left(\lambda_{2g-1}x_i^{g-1}+2\lambda_{2g}x_i^{g}+3\lambda_{2g+1}x_i^{g+1}\right) \dd x_i}{y_i} 
     -2 \dd \left(\widehat{\wp}_{ggg}\right), 
\label{6e1} 
\end{align}
where we have used $\displaystyle{\frac{y_i}{F'(x_i)}=\frac{\partial x_i}{\partial u_g}}$,\
$\displaystyle{\sum_{i=1}^g x_i=\widehat{\wp}_{gg}}$ and we denote $\widehat{\wp}_{ggg}=\wp_{ggg}/\lambda_{2g+1}$. Then we have 
\begin{align}
\dd \left(2\widehat{\wp}_{ggg}+(-\zeta_{g-1}) \right)
 &=\sum_{j=1}^g \dd u_j 
   \left(2\frac{\partial\widehat{\wp}_{ggg}}{\partial u_j}+\frac{\partial(-\zeta_{g-1})}{\partial u_j} \right)
\notag\\
 &=\sum_{j=1}^g \sum_{i=1}^g \frac{ \dd x_i x_i^{j-1}}{y_i}
   \left( 2 \widehat{\wp}_{gggj}+\lambda_{2g+1} \widehat{\wp}_{g-1,j} \right)
\notag\\
 &=\sum_{i=1}^g \frac{\left(\lambda_{2g-1}x_i^{g-1}+2\lambda_{2g} x_i^{g}+3\lambda_{2g+1} x_i^{g+1} \right) \dd x_i}{y_i},
\label{6e2} 
\end{align}
where we have used $\displaystyle{\dd u_j=\sum_{i=1}^g \frac{x_i^{j-1} \dd x_i}{y_i}}$. In the right-hand side of Eq.(\ref{6e2}), by using Eq.(\ref{2e18}), we reduce the power of $x_i$ in the range $x_i^{j-1},(j=1,2,\cdots, g)$ and comparing the coefficients of left- and right-hand side of Eq.(\ref{6e2}), we have following differential equations for general genus 
\begin{align} 
\wp_{gggj}=
 &\frac{3}{2}\wp_{gg}\wp_{gj} + \frac{3}{2}\lambda_{2g+1}\wp_{g,j-1} - \frac{1}{2}\lambda_{2g+1} \wp_{g-1,j}
\nonumber\\
 &+\lambda_{2g}\wp_{g j} + \frac{1}{2}\lambda_{2g-1}\lambda_{2g+1}\delta_{jg}, \qquad  (1 \leq j \le g) .
\label{6e3}
\end{align}
where we write differential equations with $\wp_{jk}$ and $\wp_{jklm}$ instead of $\widehat{\wp}_{jk}$, and $\widehat{\wp}_{jklm}$ to compare with our result. Then in the standard form of $\lambda_0=0$, another KdV equation 
\begin{align} 
\wp_{1 1 1, g+1-j}=
&\frac{3}{2} \wp_{11} \wp_{1, g+1-j}+\frac{3}{2} \lambda_{1} \wp_{1, g+2-j} 
-\frac{1}{2} \lambda_{1} \wp_{2, g+1-j} \nonumber\\
&+ \lambda_{2} \wp_{1, g+1-j}
+ \frac{1}{2} \lambda_1 \lambda_3 \delta_{g+1-j, 1} , \qquad (1 \leq j \le g), 
\label{6e4}
\end{align}
is satisfied for general genus.
%
%%%%%%%%%%%%%%%%%%%%%

We can obtain other differential equations for general genus recursively. For example, we start from
\begin{align}
\dd (-\zeta_{g-2})=&\sum_{i=1}^g
\frac{ \left(\lambda_{2g-3} x_i^{g-2}+2\lambda_{2g-2} x_i^{g-1}+3\lambda_{2g-1} x_i^{g}
+4\lambda_{2g} x_i^{g+1}+5\lambda_{2g+1} x_i^{g+2} \right) \dd x_i} {y_i} 
\notag\\
&-2 \dd\Big(\sum_{i=1}^g 
\frac{y_i \chi_1(x_i;x_1,x_2,\cdots, \widecheck{x}_i , \cdots,x_g}{F'(x_i)}\Big) .
\label{6e5}
\end{align}
If we notice the relation $\chi_{1}(x_i; x_1,x_2,\cdots, \widecheck{x}_i, \cdots, x_g)=\chi_{1}(x_i; x_1,x_2,\cdots, x_g)+x_i$, we have 
\begin{align}
& \sum_{i=1}^g \frac{y_i \chi_1(x_i;x_1,x_2,\cdots, \widecheck{x}_i , \cdots,x_g)}{F'(x_i)}
= \sum_{i=1}^g \left( \frac{y_i \chi_1(x_i;x_1,x_2,\cdots, x_g)}{F'(x_i)} +\frac{x_i y_i}{F'(x_i)} \right)
\notag\\
&=\sum_{i=1}^g \left( \frac{\partial x_i}{\partial u_{g-1}} + x_i  \frac{\partial x_i}{\partial u_{g}} \right)
=\frac{\partial}{\partial u_{g-1}} \left(\sum_{i=1}^{g} x_i \right)
+\frac{1}{2}\frac{\partial}{\partial u_{g} } \left(\sum_{i=1}^{g} x_i^2 \right)
\notag\\
&=\frac{\partial}{\partial u_{g-1}} \widehat{\wp}_{gg}
+\frac{1}{2}\frac{\partial}{\partial u_{g} }\left(\widehat{\wp}_{gg}^2+2 \widehat{\wp}_{g,g-1}\right)
=\widehat{\wp}_{gg}\widehat{\wp}_{ggg}+2\widehat{\wp}_{gg, g-1},
\label{6e6}
\end{align}
where we have used Eq.(\ref{2e3}).
Then we have 
\begin{align}
\dd &\left(2\widehat{\wp}_{gg}\widehat{\wp}_{ggg}+4\widehat{\wp}_{gg, g-1}+(-\zeta_{g-2}) \right)
=\sum_{j=1}^g 
\dd u_j \frac{\partial}{\partial u_j} 
\left(2\widehat{\wp}_{gg}\widehat{\wp}_{ggg}+4\widehat{\wp}_{gg, g-1}+(-\zeta_{g-2}) \right)
\notag\\
&=\sum_{i=1}^g 
\sum_{j=1}^g \frac{\dd x_i x_i^{j-1}}{y_i}
\left(2\widehat{\wp}_{ggj} \widehat{\wp}_{ggg}+2\widehat{\wp}_{gg}\widehat{\wp}_{gggj}+4\widehat{\wp}_{gg,g-1,j}
+\lambda_{2g+1} \widehat{\wp}_{g-2, j} \right)
\notag\\
&=\sum_{i=1}^g
\frac{\dd x_i}{y_i} \left(\lambda_{2g-3} x_i^{g-2}+2\lambda_{2g-2} x_i^{g-1}+3\lambda_{2g-1} x_i^{g}
+4\lambda_{2g} x_i^{g+1}+5\lambda_{2g+1} x_i^{g+2} \right) .
\label{6e7} 
\end{align}
By using Eq.(\ref{2e18}), we reduce the power of $x_i$ in the range $x_i^{j-1},(j=1,2,\cdots, g)$ and comparing coefficients of left- and right-hand side of Eq.(\ref{6e7}), we have the following differential equations for general genus
\begin{align} 
&\lambda_{2g+1} \widehat{\wp}_{g-2, j}
+2\widehat{\wp}_{g g g} \widehat{\wp}_{g g j}+2\widehat{\wp}_{g g} \widehat{\wp}_{g g g j}
+4\widehat{\wp}_{g g, g-1,j}
\nonumber\\
&=5\lambda_{2g+1} (\widehat{\wp}_{g g}^2 \widehat{\wp}_{g j} 
+\widehat{\wp}_{g, g-1}\widehat{\wp}_{g j} +\widehat{\wp}_{g g} \widehat{\wp}_{g, j-1}
+\widehat{\wp}_{g, j-2})
+4 \lambda_{2g} (\widehat{\wp}_{g g} \widehat{\wp}_{g j}+ \widehat{\wp}_{g, j-1}) \nonumber\\
&\quad+3 \lambda_{2g-1} \widehat{\wp}_{g j}+2 \lambda_{2g-2} \delta_{j, g}+\lambda_{2g-3} \delta_{j, g-1}
,\qquad ( 1\leq j \leq g)  .
\label{6e8}
\end{align}
This is another type of differential equations than Eqs.(\ref{2e34})-(\ref{2e38}), Eqs.(\ref{2e51})-(\ref{2e65}), and Eqs.(\ref{3e13})-(\ref{3e47}). 
%%%%%%%%%%%%%%%%%%%%%%%%%%%%

Another example of differential equation can be derived starting from the following equation,
\begin{equation}
\dd(-\zeta_{1})=\sum_{i=1}^g \frac{\dd x_i}{y_i}
\sum_{k=1}^{2g-1}k \lambda_{k+2}x_i^k-2 \dd \left(\sum_{i=1}^g 
\frac{y_i \chi_{g-2}(x_i;x_1, \cdots, \widecheck{x}_i , \cdots,x_g)}{F'(x_i)} \right) .
\label{6e9}
\end{equation}
If we notice relations
\begin{eqnarray}
&&\sum_{k=1}^{2g-1} \lambda_{k+2}x_i^{k+2}=y_i^2-\lambda_2 x_i^2-\lambda_1 x_i-\lambda_0, 
\nonumber\\
&&\chi_{g-2}(x_i;x_1, \cdots, \widecheck{x}_i , \cdots, x_g)
=\frac{F'(x_i)}{x_i}-(-1)^{g-1}\frac{h_{g-1}(x_1, \cdots, \widecheck{x}_i , \cdots,x_g)}{x_i}
=\frac{F'(x_i)}{x_i}-\frac{\widehat{\wp}_{g 1}}{x_i^2}, 
\nonumber
\end{eqnarray}
we have 
\begin{align}
\dd(-\zeta_{1})=&\sum_{i=1}^g 
\left( \frac{1}{x_i y_i} \dd \left(\sum_{k=1}^{2g-1}   \lambda_{k+2} x_i^{k+2} \right) 
-\frac{2}{x_i^2 y_i} \left(\sum_{k=1}^{2g-1} \lambda_{k+2} x_i^{k+2}\right) \dd x_i \right)
\notag\\
&-2 \dd \left(\sum_{i=1}^g 
\left( \frac{y_i}{x_i} -\frac{ y_i \widehat{\wp}_{g 1}}{x_i^2 F'(x_i)}
\right) \right)
\notag\\
=&\sum_{i=1}^g \left( \frac{1}{x_i y_i}
\dd \left(y_i^2-\lambda_2 x_i^2-\lambda_1 x_i\right) -\frac{2}{x_i^2 y_i}
\left(y_i^2-\lambda_2 x_i^2-\lambda_1 x_i -\lambda_0\right)\dd x_i -2\dd \left(\frac{ y_i}{x_i} \right)\right)
\notag\\
&+2\dd \left( \widehat{\wp}_{g 1}  \sum_{i=1}^g \frac{1}{x_i^2} \frac{\partial x_i}{\partial u_g}
\right)
\notag\\
=&\sum_{i=1}^g \frac{\dd x_i}{y_i}\left(\frac{\lambda_1}{x_i}+\frac{2\lambda_0}{x_i^2}\right)
-2 \dd \left( \widehat{\wp}_{g 1} \frac{\partial }{\partial u_g} \left( \sum_{i=1}^g \frac{1}{x_i}\right)\right)
\notag\\
=&\sum_{i=1}^g \frac{\dd x_i}{y_i}\left(\frac{\lambda_1}{x_i}+\frac{2\lambda_0}{x_i^2}\right)
+2 \dd \left( \widehat{\wp}_{g 1} \frac{\partial }{\partial u_g} 
\left(\frac{\widehat{\wp}_{g 2} }{ \widehat{\wp}_{g 1} }\right)\right)
\notag\\
=&\sum_{i=1}^g \frac{\dd x_i}{y_i}\left(\frac{\lambda_1}{x_i}+\frac{2\lambda_0}{x_i^2}\right)
+2 \dd \left(\widehat{\wp}_{g g 2}-\frac{\widehat{\wp}_{g 2} \widehat{\wp}_{g g 1} }{\widehat{\wp}_{g 1} } \right) ,
\label{6e10}
\end{align}
where we have used $\displaystyle{\sum_{i=1}^g \frac{1}{x_i}=-\frac{\widehat{\wp}_{g 2}}{\widehat{\wp}_{g 1}} }$. Then we have  
\begin{align}
\dd &\left(-2\widehat{\wp}_{g g 2}+2\frac{\widehat{\wp}_{g 2} \widehat{\wp}_{g g 1} }{\widehat{\wp}_{g 1} } 
+(-\zeta_{1}) \right)
=\sum_{i=1}^g \sum_{j=1}^g 
\frac{\dd x_i x_i^{j-1}}{y_i}\frac{\partial }{\partial u_j}
\left(-2\widehat{\wp}_{g g 2}+2\frac{\widehat{\wp}_{g 2} \widehat{\wp}_{g g 1} }{\widehat{\wp}_{g 1} } 
+(-\zeta_{1}) \right)
\notag\\
&=\sum_{i=1}^g \sum_{j=1}^g 
\frac{\dd x_i x_i^{j-1}}{y_i}
\left(-2\widehat{\wp}_{g g 2 j}+2\frac{\widehat{\wp}_{g 2 j } \widehat{\wp}_{g g 1} }{\widehat{\wp}_{g 1} } 
+2\frac{\widehat{\wp}_{g 2 } \widehat{\wp}_{g g 1 j} }{\widehat{\wp}_{g 1} } 
-2\frac{\widehat{\wp}_{g 2 } \widehat{\wp}_{g g 1} \widehat{\wp}_{g 1 j}}{\widehat{\wp}_{g 1} ^2} 
+\lambda_{2g+1} \widehat{\wp}_{1 j}
\right)
\notag\\
&=\sum_{i=1}^g \frac{\dd x_i}{y_i}\left(\frac{\lambda_1}{x_i}+\frac{2\lambda_0}{x_i^2}\right) .
\label{6e11}
\end{align}
While, using Eq.(\ref{2e18}) in the form
$$
\frac{1}{x_i}=\frac{1}{\widehat{\wp}_{g1}}  x_i^{g-1}-\frac{\widehat{\wp}_{gg}} {\widehat{\wp}_{g1}} x_i^{g-2}
-\frac{\widehat{\wp}_{g,g-1}}{\widehat{\wp}_{g1}} x_i^{g-3}-
\cdots -\frac{\widehat{\wp}_{g2}}{\widehat{\wp}_{g1}},
$$
puting the power of $x_i$ in 
the range $x_i^{j-1},(j=1,2,\cdots, g)$ and comparing coefficients of left- and right-hand side of Eq.(\ref{6e11}), we have following differential equations for general genus
\begin{align} 
&\lambda_{2g+1}  \widehat{\wp}_{1 j}-2 \widehat{\wp}_{g g 2 j}
+\frac{2 \widehat{\wp}_{g 2  }\widehat{\wp}_{g g 1 j}}{ \widehat{\wp}_{g 1}}
+\frac{2 \widehat{\wp}_{g g 1}\widehat{\wp}_{g 2 j  }}{ \widehat{\wp}_{g 1}}
-\frac{2 \widehat{\wp}_{g 2  }\widehat{\wp}_{g g 1   }  \widehat{\wp}_{g 1 j}}{\widehat{\wp}_{g 1}^2}
\nonumber 
\\
&=-\frac{ \lambda_1  \widehat{\wp}_{g,j+1}}{\widehat{\wp}_{g 1}} 
  -\frac{2\lambda_0  \widehat{\wp}_{g,j+2}}{\widehat{\wp}_{g 1}}
  +\frac{2\lambda_0  \widehat{\wp}_{g 2}    \widehat{\wp}_{g,j+1}}{\widehat{\wp}_{g 1}^2}
  +\frac{2\lambda_0}{\widehat{\wp}_{g 1}} \delta_{j,g-1}
  +\frac{ \lambda_1}{\widehat{\wp}_{g 1}} \delta_{jg} 
  -\frac{2\lambda_0  \widehat{\wp}_{g 2}} {\widehat{\wp}_{g 1}^2} \delta_{jg} , 
\quad ( 1\leq j \leq g) .
\label{6e12}
\end{align}
We have explicitly checked Eq.(\ref{6e8}) and Eq.(\ref{6e12}) for $g=3$ and $j=1, 2, 3$.
}
%%%%%%%%%%%%%%%%%%%%%%%%%%%%%%%%%%%%%%%%%%%%%%%%%%%%%%%%%%%%%%%%%%%%
%%%%%%%%%%%%%%%%%%%%%%%%%%%%%%%%%%%%%%%%%%%%%%%%%%%%%%%%%%%%%%%%%%%%%%
\subsection{Hirota form differential equations} 

For genus two case, all differential equations Eqs.(\ref{2e34})-(\ref{2e38}) are written in the Hirota form, that is, bilinear differential equation with Hirota derivative. For genus three and four cases, though the left hand side can be written in the Hirota form, but differential equations which contain $\Delta$  are not written in the Hirota form. As it is quite natural, Baker already has used the Hirota derivative for the derivative of $(-\log \sigma)$, that is, $\wp_{jk}$ and $\wp_{jklm}$~\cite{Baker3}. We use following relations 
\begin{align}
(\log \tau )_{xy}=&\frac{D_x D_y \tau \cdot \tau}{2 \tau^2}  , 
\label{4e11}\\
(\log \tau )_{xyzt}=&\frac{D_x D_y D_z D_t \tau \cdot \tau}{2 \tau^2}
 -\frac{(D_x D_y  \tau \cdot \tau) (D_z D_t  \tau \cdot \tau) }{2 \tau^4}
 -\frac{(D_x D_z  \tau \cdot \tau) (D_y D_t  \tau \cdot \tau) }{2 \tau^4}\nonumber\\
&-\frac{(D_x D_t  \tau \cdot \tau) (D_y D_z  \tau \cdot \tau) }{2 \tau^4}  ,
\label{4e12}
\end{align}
where $D_x$, $D_y$, $D_z$ and $D_t$ are Hirota derivatives. Just as the Weierstrass $\wp$ function solution in the KdV equation, we identify the $\tau$ function in such a way as $(-\log \tau)$ is proportional to $(-\log \sigma)$~\cite{Hayashi4}. Then we put
$\displaystyle{ \wp_{jk}=(-\log \sigma)_{jk}=\alpha (-\log \tau)_{jk}}$
with constant $\alpha$. We show that
$\displaystyle{I=\tau^2\Big(\wp_{xyzt}-\frac{1}{2}(\wp_{xy} \wp_{zt}+\wp_{xz} \wp_{yt}+\wp_{xt} \wp_{yz}) \Big)}$
can be written in the Hirota form in the following way
\begin{align}
I=&\tau^2 \Big( \wp_{xyzt}-\frac{1}{2}(\wp_{xy} \wp_{zt}+\wp_{xz} \wp_{yt}+\wp_{xt} \wp_{yz})\Big)
\nonumber\\
=& (-\alpha) \tau^2 \Big[
(\log \tau )_{xyzt}+\frac{\alpha}{2} \Big((\log \tau )_{xy} (\log \tau )_{zt}+(\log \tau )_{xz} (\log \tau )_{yt}
+(\log \tau )_{xt} (\log \tau )_{yz}\Big) \Big] \nonumber\\
=&-\frac{\alpha}{2} 
\bigg[ 
D_x D_y D_z D_t \tau \cdot \tau
-\left(1-\frac{\alpha}{4}\right)\left(\frac{(D_x D_y  \tau \cdot \tau) (D_z D_t  \tau \cdot \tau) }{\tau^2}
+\frac{(D_x D_z  \tau \cdot \tau) (D_y D_t  \tau \cdot \tau) }{\tau^2}\right.\nonumber\\
&
\hspace*{60mm}\left.
+\frac{(D_x D_t  \tau \cdot \tau) (D_y D_z  \tau \cdot \tau) }{\tau^2}\right)\bigg]
\nonumber 
\\
=&-2D_x D_y D_z D_t \tau \cdot \tau .
\label{4e13}
\end{align}
where in the last step we choose $\alpha=4$. For more general form, we have  
\begin{align}
J=&\tau^2 \left( \wp_{xyzt}-\frac{1}{2}(\wp_{xy}\wp_{zt}+\wp_{xz}\wp_{yt}+\wp_{xt}\wp_{yz})
+a \wp_{xy} +b \right) \nonumber\\
=&-2\left(D_x D_y D_z D_t \tau \cdot \tau +a D_x D_y \tau \cdot \tau
-\frac{1}{2} b \tau^2 \right)  .
\label{4e14}
\end{align}
The l.h.s. of Eqs.(\ref{2e34})-(\ref{2e38}), Eqs.(\ref{2e51})-(\ref{2e65}), and Eqs.(\ref{3e13})-(\ref{3e47}) and the linear term of $\wp_{jk}$ and constant term in the r.h.s. can be written in the generalized Hirota form, which contains $\text{(const.)} \times \tau^2$ term, such as the Hirota form for Weierstrass $\wp$ solution in the KdV equation~\cite{Hayashi4}. Equations which contain $\Delta$ and $\Delta_i$ cannot be written as the Hirota bilinear differential form.

%
%%%%%%%%%%%%%%%%%%%%%%%%%%%%%%%%%%%%%%%%%%%%%%%%%%%%%%%%
\section{Summary and Discussions} 
\setcounter{equation}{0}

In order to find higher dimensional integrable models, we have explicitly studied 
how to obtain differential equations of genus four hyperelliptic $\wp$ function.

In the standard form of $\lambda_0=0$, 
we have KdV and another KdV equations for genus being more than two.
In the standard form of $\lambda_1=0$, if genus is three, we have KP equation.
The universality of integrable model is guaranteed up to three dimensional 
integrable models. 
As the two- and three-dimensional integrable models, KdV equation and 
KP equation come out, respectively. 

If genus is two, all differential equations are written in the Hirota 
form. 
However, we obtain differential equations which cannot 
be written in the Hirota form, if genus is more than three. 
This means that the Hirota form or the fermionic bilinear form is 
not sufficient to characterize higher dimensional integrable models.

From the series of investigations of 
genus two, three, and four, differential equations for general genus will not be 
so complicated, but only the quadratic term $\Delta_j$ 
of $\wp_{jk}$ becomes complicated. 

We have also shown, in the standard form of $\lambda_0=0$, some duality for 
the set of differential equations, which gives that 
KdV and another KdV equations always exist for genus being more than two. 
In the standard form of $\lambda_1=0$, there also exist duality for the KP equation 
for genus three and four. 
We expect that the same expression Eq.(\ref{2e65}) and/or Eq.(\ref{3e47}) 
will be satisfied for the general genus. 
 
Since we have KdV, another KdV equation, and $(g-2)$ pieces of KP type differtial equations in the standard form of $\lambda_0=0$, where KP type  equations are similar to the KdV equation, we expect that genus $g$ hyperelliptic $\wp$ functions have rank $g$ Lie group structure. 
%
%%%%%%%%%%%%%
Actually, hyperelliptic $\wp$ functions can be defined through hyperelliptic $\vartheta$ functions instead of hyperelliptic curves. For the genus two case, hyperelliptic theta functions are written in the form 
$\vartheta\begin{bmatrix} a & c \\ b & d \end{bmatrix}(u_1, u_2;\tau_{11}, \tau_{22}, \tau_{12})$, where $a,b,c,d=0\ \text{or}\ 1/2$.
If we take the special limit of $\tau_{12} \rightarrow 0$, 
 they reduce to the product of genus one theta functions in the form
$\vartheta\begin{bmatrix} a & c \\ b & d \end{bmatrix}(u_1, u_2;\tau_{11}, \tau_{22}, 0)%
=\vartheta\begin{bmatrix} a \\ b \\ \end{bmatrix}(u_1;\tau_{11})%
 \vartheta\begin{bmatrix} c \\ d \\ \end{bmatrix}(u_2;\tau_{22})$. %
In this special limit, we have two independent one dimensional KdV equations (static KdV equations), which correspond to Eq.(\ref{2e34}) and Eq.(\ref{2e38}). As we know that there is an SO(2,1) structure in each one dimensional KdV equation~\cite{Hayashi2}, we have \mbox{SO(2,1) $\otimes$ SO(2,1)} structure in the genus two hyperelliptic $\wp$ functions. For the general genus cases, the situation is the same and we expect that there is rank $g$ Lie grroup structure for the genus $g$ hyperelliptic $\wp$ functions. 

In some special cases, we have given some differential equations for general genus by using our method step by step. 
The differential equations of hyperelliptic $\wp$ functions are integrable, then we expect that Lax pairs for these differential equations exist. Parshin have succeeded in generalizing the KP hierarchy with the Lax pair~\cite{Parshin}, but it is the genus 0 case, that is, the solution is expressed by polynomial, exponential and trigonometric functions. The Lax pair of genus two will contain three potentials, $\wp_{22}(u_1, u_2)$, $\wp_{21}(u_1, u_2)$, and $\wp_{11}(u_1, u_2)$, will give the set of 5 equations but not the single KP equation, so that it seems difficult to construct the Lax pair in this case.

\newpage
%%%%%%%%%%%%%%%%%%%%%%%%%%%%%%%%%%%%%%%%%%%%%%%%%%%%%%%%%%%
%%%%%%%%%%%%%%%%%%%%%%%%%%%%%%%%%%%%%%%%%%%%%%%%%%%%%%%%%%%  

\end{document}